\countdef\FIGS=26
\countdef\TWOCOL=23

\ifnum\TWOCOL=1
\documentstyle[aps,twocolumn,prb,epsfig]{revtex}
\else
\documentstyle[aps,prb]{revtex}
\fi

\newcommand{\etal}{{\it et al.}}
\newcommand{\ie}{{\it i.e.}}
\newcommand{\eg}{{\it e.g.}}

\newcommand{\phiext}{\Phi_{\mathrm{ext}}}
\newcommand{\phitot}{\Phi_{\mathrm{tot}}}
\newcommand{\phiscr}{\Phi_{\mathrm{scr}}}
\newcommand{\phioff}{\Phi_{\mathrm{off}}}
\newcommand{\betal}{\beta_\ell}

\newcommand{\InLineFrac}[2]{\frac{#1}{#2}}

\newcommand{\Fig}[1]{Fig.~\ref{#1}}
\newcommand{\MultFig}[2]{\Fig{#1}(#2)}

\newcommand{\InLineRef}[1]{Ref.~\onlinecite{#1}}

\newcommand{\MagImageHistCaption}{
(a) Representative paramagnetic magnetization image for the array,
with an external cooling field of $4.8 \Phi_0$. The color scale 
runs from $-0.6 \Phi_0$ (white) to 
$0.6\Phi_0$ (black). (b) Magnetization
histogram generated from the paramagnetic image. 
}

\newcommand{\MagVsFluxCaption}{
Measured mean magnetization of the array  
$\langle\phitot - \phiext\rangle 
/ \Phi_0$, plotted versus cooling field, $\phiext/
\Phi_0$. The error for each data point is less than the size of the 
plotted symbols.}

\newcommand{\ScreeningImageCaption}{Mechanisms for generating
the diamagnetic screening current around a multiply connected 
superconducting sample: (a) The conventional picture,  where each loop
of the sample generates a current. (b) Our new picture,
where only the exterior plaquettes create a screening current. This
creates a diamagnetic current on the outside of the sample,
and a paramagnetic current just inside the sample.
}

\begin{document}

\draft

\author{A. P. Nielsen \cite{nielsen:addr} \and
        A. B. Cawthorne \cite{cawthorne:addr}\and 
        P. Barbara  \and 
        F. C. Wellstood \and
        C. J. Lobb}
\address{Center for Superconductivity Research, University of Maryland,
College Park, MD 20742}

\author{R. S. Newrock}
\address{Department of Physics, University of Cincinnati,
Cincinnati, OH 45421}

\author{M. G. Forrester}
\address{Northrop Grumman Corporation, 1310 Beulah Rd, Pittsburgh, PA  15235}

\title{Paramagnetic Meissner Effect in Multiply-Connected Superconductors}
\date{\today}

\ifnum\TWOCOL=1
\twocolumn[\hsize\textwidth\columnwidth\hsize\csname@twocolumnfalse\endcsname
\fi

\maketitle

\begin{abstract}
We have measured a paramagnetic Meissner effect in 
$\mathrm{Nb}-\mathrm{Al}_2\mathrm{O}_3-\mathrm{Nb}$ 
Josephson junction arrays using a scanning SQUID microscope.
The arrays exhibit diamagnetism for some cooling fields and
paramagnetism for other cooling fields. The measured
mean magnetization is always small, 
$\langle\phitot-\phiext\rangle 
< 0.3 \Phi_0$
(in terms of flux per unit cell of the array, where $\Phi_0$ is the
flux quantum)
for the range of cooling fields investigated ($-12\Phi_0$ to $12 \Phi_0$).  
We demonstrate that a new model of magnetic screening, 
valid for multiply-connected 
superconductors, reproduces all of the essential features of 
paramagnetism that we observe and that no exotic mechanism, such 
as $d$-wave superconductivity, is needed for paramagnetism. 
\end{abstract}


\ifnum\TWOCOL=1
]\fi

A paramagnetic response was first observed in BSCCO
\cite{braunisch_papers} and subsequently reported in many other
high $T_c$ materials.\cite{schliepe1,riedling1,onbasli1,okram1}
This response is in striking contrast to the diamagnetic response
which is a hallmark of superconductivity. 
It has been argued that the 
paramagnetic Meissner effect (PME)  results
from a $d$-wave symmetry in the
BSSCO order parameter.\cite{sigrist1}
The observation of PME in 
niobium\cite{thompson1,kostic1,terentiev1} and aluminum,\cite{geim1} 
known to be $s$-wave superconductors,
questioned the validity of the $d$-wave argument.

There are several proposed theories
for the  PME,
although no experiment has been yet able to provide 
conclusive evidence for one or another; and there is  
disagreement as to the validity of the $d$-wave 
explanation.\cite{note:disagree}
In high $T_c$ materials one possible
explanation is the existence of $\pi$-junctions which result
from the misalignment of grains in materials with $d$-wave symmetry.
\cite{sigrist1} 
In the conventional
materials, which are known to be $s$-wave, PME has been attributed to 
different mechanisms causing non-equilibrium flux
configurations. A non-equilibrium
flux state may result from flux-compression,\cite{koshelev1} the
similar giant vortex state,\cite{moshchalkov1,deo1}
pinning randomness, or a surface barrier at the
edge of the sample.\cite{deo2} 

To create a controlled experiment we have chosen to  look at 
Josephson-junction arrays using a scanning SQUID microscope (SSM),
using a technique similar to that of
Kirtley {\it et al.}\cite{kirtley1} The SSM provides 
spatially resolved magnetization images of the sample. 
Arrays eliminate
many of the complications in studying the underlying cause of the PME. 
Since our arrays are lithographically fabricated from $\mathrm{Nb}$ 
and $\mathrm{Al}_2\mathrm{O}_3$,
we know that there are neither
$\pi$-junctions nor significant randomness.
Additionally, our sample is uniformly attached to the thermal bath,
eliminating thermal gradients across the sample during cooling. 
We find that arrays can be either paramagnetic or diamagnetic 
depending upon the cooling field. Additionally, 
we  propose a new model, valid
for multiply-connected supercondutors, which predicts a paramagnetic
response and agrees with our scanning SQUID measurements. 

The arrays consist of a square array of niobium crosses in two layers,
with a unit cell size of $a=46 \mu \mathrm{m}$.
Junctions are formed at the cross overlap.
The calculated  self-inductance of each loop of four junctions  is
$L=64 \mathrm{pH}$ 
and the measured critical current density is 
$J_c = 600 \mathrm{A}/\mathrm{cm}^2$ at $4.2\mathrm{K}$, with a 
junction area of $ 5 \times 5 \mu \mathrm{m}^2$. The
entire array consists of $30 \times 100$ junctions. 
The unit-cell dimensions
compare to the typical grain size seen in BSCCO samples
\cite{braunisch_papers,kirtley1} which exhibit PME, to which similar
fields are applied. 
The array is placed 
in an 
SSM sample stage,\cite{black2} 
sitting at the end of a sapphire rod of $10 \mathrm{mm}$ 
diameter,  around which
a solenoid has been wound.  Before any data is collected we first 
remove any flux trapped within the pick up loop of the SQUID
to avoid interaction with the sample.\cite{mathai2} Additionally, we measure 
and correct for any background field present.

We cooled our arrays to 4.2K in fixed external magnetic field, 
and then used the SSM to measure the magnetic flux threading the 
SQUID from the sample, $\phitot$, as a function
of position. 
Our spatial resolution was limited by our SQUID-sample separation
distance, which we estimate to be between $40$ and $ 60 \mu 
\mathrm{m}$. 
Data were taken every $5 \mu \mathrm{m} $ in the 
$x$ direction and 
every $50 \mu \mathrm{m}$ in the $y$ direction, and converted from flux
through the SQUID to
normalized flux per unit cell. We measure the external flux, $\phiext$, 
directly by 
warming the sample above $T_c$ and imaging. 
From the data we calculate both a local normalized magnetization
$(\phitot - \phiext) / 
\Phi_0$ and an average overall 
magnetization
$\langle\phitot - \phiext\rangle / 
\Phi_0$.\cite{units:note}
We were unable to resolve individual unit cells in the array,
\cite{samplesize:note} but we were still able to resolve
large scale flux distributions.

Our measurement and analysis proceeded in a fashion
similar
to \InLineRef{kirtley1}.  
We first make an image of the magnetization
of the array (a typical 
image is shown in \MultFig{magpicts}{a}). 
We then make a magnetization histogram and compute the mean
magnetization. \MultFig{magpicts}{b} shows the 
magnetization histogram generated from the image in
\MultFig{magpicts}{a}. The histogram is computed
by looking at the lower third of the array
and counting the number of points at a
given magnetization.\cite{scratches:note} We find that the array 
may be either diamagnetic or paramagnetic, but surprisingly, appears
to be preferentially paramagnetic. Additionally, we always observe
diamagnetic screening currents around the outside edges of the sample. 
The array shown in \MultFig{magpicts}{a} was cooled in an external
field of $ 4.8 \Phi_0$ and 
clearly is paramagnetic ({\it i.e.} it is darker than the 
background). We find a mean 
paramagnetic response of 
$\langle\phitot-\phiext\rangle 
= (0.063 \pm 0.005) \Phi_0$ for this cooling field.

\ifnum\FIGS=1
\begin{figure}[h]
\epsfig{file=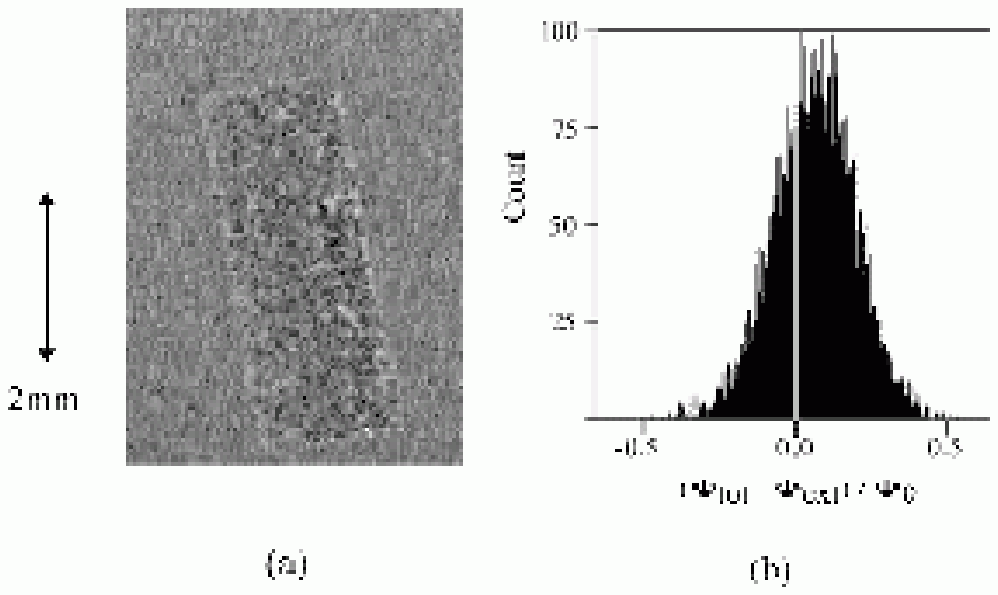,width = 3.25in, angle =0}
\caption{\MagImageHistCaption }
\label{magpicts}
\end{figure}
\fi


The data plotted in \Fig{mvh:data} 
show
the mean magnetization versus external cooling field. 
Each data point represents a single field-cooled average 
measurement. The error in the mean values plotted, such as discussed 
above, is less than the size of each data point as drawn.
The sign
of the magnetization is reproducible, and the magnitude is
reproducible to within a factor of two
when the sample is warmed above 10K and recooled in
the same field, indicating that our external field is consistent
to within $ \pm 0.05 \Phi_0$ 
The external field has a spatial variation of $\pm 0.5 \Phi_0$, 
which is limited by the gradient of the background field in 
our probe. 
Although there is some variation from one cooling
to the next in the actual magnetic images for the same external field,
the paramagnetism is remarkably reproducible.  

\ifnum\FIGS=1
\begin{figure}[h]
\epsfig{file=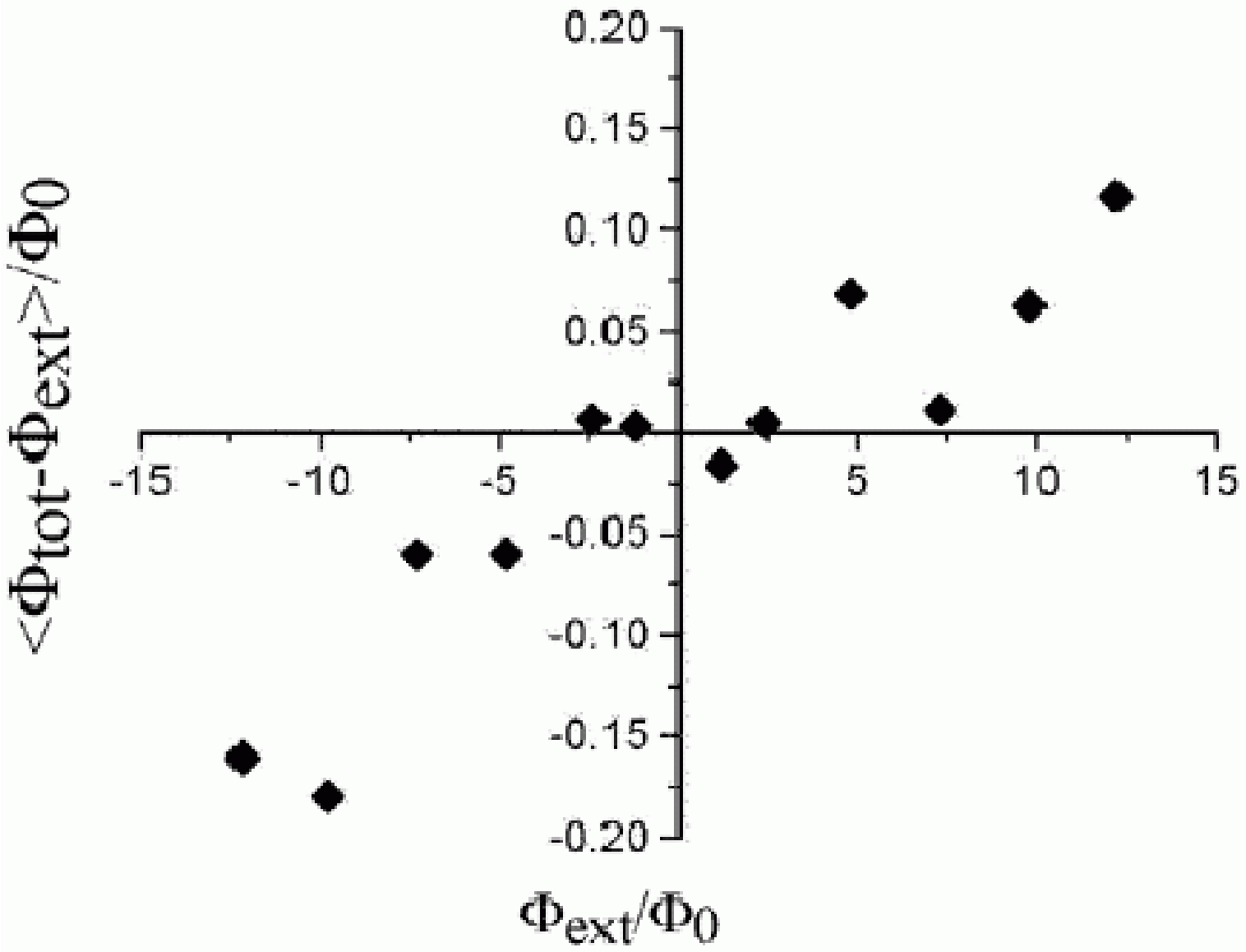,width=3.0in,angle=0}
\caption{\MagVsFluxCaption}
\label{mvh:data}
\end{figure}
\fi


We may consider a single loop of the array as a 
jumping off point for modeling
the entire array's magnetization. 
Originally used
to describe the observed magnetization of a single 
loop with a single Josephson-junction,\cite{silver1} 
Sigrist and Rice
\cite{sigrist1} later used this model to argue that
$\pi$-junctions could cause PME in $d$-wave superconductors. 
The single-loop model
gives a relationship between the external flux 
applied to a four-junction loop and the total measured flux
\begin{equation}
\phitot = \phiext + 
L I_c \sin\left({\pi\over 2} n 
               - {\pi\over 2} {\phitot \over \Phi_0}\right).
\label{phitotvsext}
\end{equation}
Here $I_c$ is the critical current of a single junction.
For all values of $\InLineFrac{L I_c}{\Phi_0} >\InLineFrac{2}{\pi}$, 
$\phitot$ is a
multi-valued function of $\phiext$ depending on an
integer $n$. 
We note here that \InLineRef{sigrist1} used a model with a 
$\InLineFrac{L I_c}{ \Phi_0}$ slightly greater than one to 
demonstrate the absence of potential 
paramagnetism
for a single loop with a single junction. Had they used a value of
$\InLineFrac{L I_c}{ \Phi_0} \gg 1$ 
they would have noticed the potential
for paramagnetism even in the absence of external flux, although
such a situation is not energetically favorable. 
For a four junction loop,
$\betal = 2 \pi \InLineFrac{L I_c}{ \Phi_0} = 30$
allows for the possibility of paramagnetism (or diamagnetism) 
at $\phiext=0$.

Computing the Gibbs free energy
for the various
possible solutions to (\ref{phitotvsext}) we find that for
$m<\InLineFrac{\phiext}{\Phi_0} <m+\frac{1}{2}$ the single loop will
be diamagnetic and for $m+\frac{1}{2}<\InLineFrac{\phiext}{\Phi_0} <m+1$
will be paramagnetic, where $m$ is any integer. 
If one applies a positive offset flux $\phioff$ to the 
four-junction loop (\ie\ the total applied flux is 
$\phiext + \phioff$) the solutions $\phitot - \phiext$ {\it vs.} $\phiext$ 
become
more paramagnetic and conversely if $\phioff < 0$ the solutions 
become more diamagnetic. 
If, \eg, a 
simple
diamagnetic screening current flows around the outside edge of the
superconductor (negative offset), the measured magnetization will be more
diamagnetic.


A conventional superconductor is  a perfect diamagnet:
An applied flux causes screening currents to flow around the outside. 
In an ideal array of Josephson junctions, one may
decompose this edge current, $I$, into loop currents of magnitude $I$,
flowing around each plaquette, as shown in
\MultFig{screeningfig}{a}.
In a real
array, this is not likely to be the complete picture. 
The currents will flow only
within one penetration depth in the superconductor. For our array, 
the penetration depth is $900 \mathrm{\AA}$ while the wire size is 
$10 \mu \mathrm{m}$ so that the interior loop currents 
do not cancel. This means that
the array must create an energy $\frac{1}{2} LI^2$ for each plaquette in order
to generate a screening current $I$ for the array. 
In an overlap geometry, the
currents through the Josephson-junctions are uniform, so the currents
through the junctions still cancel and we do not have any increase in the
total Josephson energy, $E_J$. For an array of
$N \times N$ junctions \MultFig{screeningfig}{a} costs a 
total energy of 
\begin{equation}
E_a = \frac{1}{2} L I^2 N^2 + 4 N (1 - \cos \gamma) E_J
\label{eq:conventionalenergy}
\end{equation}
where $\gamma$ is the gauge invariant phase difference. 

\ifnum\FIGS=1
\begin{figure}[h]
\epsfig{file=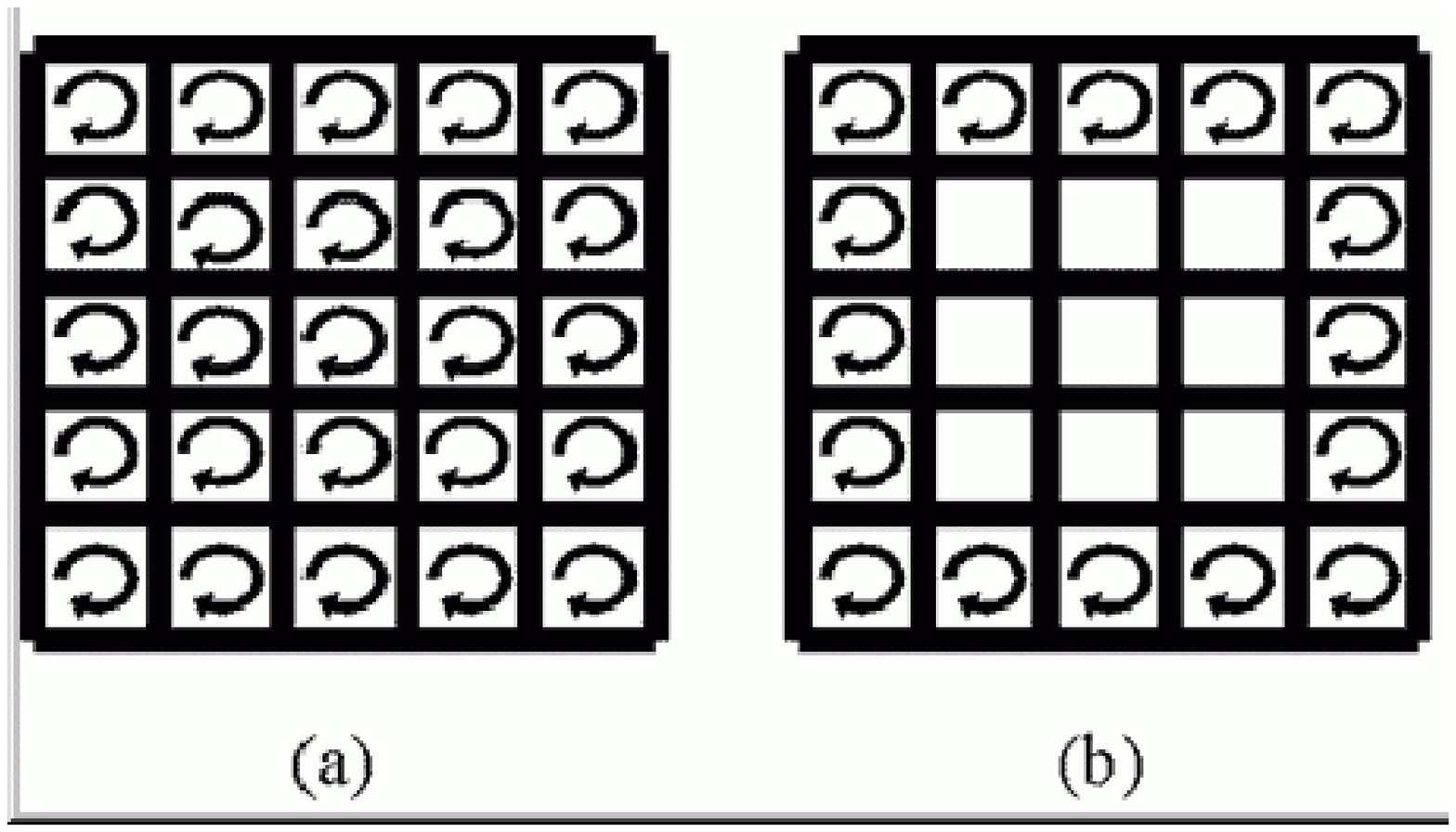,width = 3.0in}
\caption{\ScreeningImageCaption }
\label{screeningfig}
\end{figure}
\fi


We propose that the array actually screens in a much different fashion. 
The array may generate 
a diamagnetic screening current around the outside perimeter
if each  
exterior plaquette maintains  a screening current. as
shown schematically in \MultFig{screeningfig}{b}. 
In addition to the diamagnetic screening current, we then 
also have a paramagnetic current, of the same magnitude,  
flowing just inside the array. 
The array must generate the $\frac{1}{2} LI^2$ energy only for
each exterior plaquette, so the energy cost to the array in this 
case is
\begin{equation}
E_b = \frac{1}{2} L I^2 (4N) + \lbrace 4N+4(N-1) \rbrace (1-\cos\gamma)E_J.
\label{eq:newenergy}
\end{equation}
It is clear from (\ref{eq:conventionalenergy}) and
(\ref{eq:newenergy}) that $E_a \propto N^2$ while
$E_b \propto N$ so that for a large enough array, 
the conventional picture of screening will \emph{not}
be energetically favorable. 
\MultFig{screeningfig}{b} has lower energy than 
\MultFig{screeningfig}{a} when, 
assuming $\gamma$ is proportional to 
$\phitot$,
\begin{equation}
\betal > {4N-1 \over N^2-4N}, \ \forall \ N>2.
\label{eq:criteria}
\end{equation}
For our array the criteria is easily 
satisfied and we expect screening to take place as in 
\MultFig{screeningfig}{b}. 


In \MultFig{screeningfig}{a} we can treat the outer ring of
Josephson-junctions as a single loop of $4N$ junctions. 
This loop will screen up to $\InLineFrac{LI_c}{\Phi_0}=4.8$ flux quanta 
from the interior of the array, when $I_c$ is flowing around
the exterior of the array. In \MultFig{screeningfig}{b}
each exterior plaquette can screen up to $\InLineFrac{LI_c}{\Phi_0}$ flux
quanta \emph{from itself}.
This is a much larger absolute field
than in the conventional case and the difference can be easily
observed experimentally. We have observed, that flux
will not penetrate the array until the external flux exceeds 
at least
$4.0 \Phi_0$ per unit cell. Additionally, ac susceptibility
work demonstrated
that little effect was seen until the amplitude of the 
driving ac flux exceeded $4.6 \Phi_0$ per unit cell of the 
array.\cite{barbara_ac_papers}


The 
interior plaquettes see the external field, 
a diamagnetic flux from the extreme outside
loop of the array, and a paramagnetic flux from  
just inside the array. Because the loop generating the paramagnetic
flux is closer, the interior loops will see a net paramagnetic flux in 
addition to the external field.
Considered individually, 
the interior plaquettes obey a modified Eq.~(\ref{phitotvsext}),
\begin{equation}
\phitot = \phiext + \phiscr +
L I_c \sin\left({\pi\over 2} n 
               - {\pi\over 2} {\phitot \over \Phi_0}\right)
\label{eq:phitotvsextmod}
\end{equation}
where $\phiscr$ is the  flux seen by the inner plaquette due
to the screening currents produced by the exterior plaquettes.
When $\phiscr > 0$ 
the interior plaquette sees a paramagnetic offset  and its magnetization
will be more paramagnetic. 
This causes the array
to be paramagnetic more often than diamagnetic in agreement with
experiment.


If we consider only the mutual inductance between a test
plaquette in the interior of the array and the exterior plaquettes
which provide the screening, we can make a crude estimate
of the paramagnetic response. 
The maximum current that the junctions can sustain is the critical
current, $I_c$; this sets the upper bound for the magnetization 
offset. We have numerically computed the flux induced from the 
screening currents in \MultFig{screeningfig}{b}
onto the interior plaquettes
for our $30 \times 100$ junction array. 
When the screening currents are $I_c$, the maximum flux is
induced, and the \emph{minimum} $\phiscr$ we compute 
is $0.15 \Phi_0$ per unit cell, near the center of the
array. 
This means that the center plaquette should have its magnetization
shifted upward by $0.15 \Phi_0$. 
Plaquettes nearer to the exterior see 
$\phiscr$ values as large as several $\Phi_0$. 
Additionally,
we compute an average flux induced into the interior 
of the array to be $\phiscr = 0.277\Phi_0$ per unit cell. 
Considering the crudeness of the model, this is remarkable agreement
with our measured magnetization (see \Fig{mvh:data}).

We expect that the screening current  will 
increase monotonically with the external flux, up to $I_c$, and therefore
expect that the paramagnetic
offset in the interior plaquette magnetization will increase as 
the external flux increases. This is also consistent with our observations 
in Fig.~\ref{mvh:data}.


The model described here does not account for many features
of the real array. We have not accounted for variations in the
Josephson-junction parameters and we have
ignored the mutual induction  between interior plaquettes.   
Phillips \etal\cite{phillips1} showed the
importance of including the mutual inductance matrix when computing
array magnetization.
Chandran has, in the overdamped junction limit, used a model
including the full inductance matrix and computed a paramagnetic
moment in a Josephson-junction array.\cite{chandran1}
Unfortunately, our arrays are highly 
underdamped, which makes a model utilizing the full-inductance
and dynamics much more computationally difficult.
\cite{note:meanfield}
The simplicity of our model  does not describe all of the features
that we observe in our array. The observed flux pattern
is very complicated, with both paramagnetic and diamagnetic
regions.

We have shown that in a multiply connected superconductor such 
as our array, that the lowest energy configuration that provides
diamagnetic screening will also tend to make the interior of the
array paramagnetic. Furthermore, the predictions of this model
are consistent with our measurements, that the array is more often
paramagnetic than diamagnetic.
We believe that the paramagnetism in Josephson-junction 
arrays is best described by the effects detailed above. 

We would like to thank J. R. Kirtley, M. Sigrist and M. Chandran
for useful communications.
Additionally we would like to thank R. L. Greene for his critical reading
of our manuscript.  
We acknowledge
support from 
AFOSR Grant F49620-98-1-0072, NSF Grant DMR-9732800 and 
NSF Grant DMR-9801825.

\ifnum\FIGS=0

\begin{figure}
\caption{\MagImageHistCaption }
\label{magpicts}
\end{figure}

\begin{figure}
\caption{\MagVsFluxCaption}
\label{mvh:data}
\end{figure}

\begin{figure}
\caption{\ScreeningImageCaption}
\label{screeningfig}
\end{figure}

\fi


\begin{thebibliography}{10}
\bibitem{units:note}
$\phiext = H a^2 $ and $\phitot = B a^2 $ so that $(\phitot-\phiext)=4 \pi M
  a^2 $ where $H=B-4 \pi M$ and $B$ is the magnetic induction, $H$ is the
  magnetic field and $M$ is the magnetization, in Gaussian units.
\bibitem{samplesize:note}
The size of our SQUID is $10 \times 10 \mu \mathrm{m}^2$, while the separation
  between the SQUID and sample is between $40$ and $60 \mu \mathrm{m}$. The
  SQUID-sample separation limits the spatial resolution we can achieve to
  roughly $50 \mu \mathrm{m}$. Because of this, and the design of our sample
  stage, it was not feasible to scan at a higher than $50 \mu \mathrm{m}$
  resolution in the $y$ direction. Because we do not undersample the array the
  average we quote for the magnetization of the array will be correct.
\bibitem{scratches:note}
Scratches were accidentally introduced into the array {\it in situ}. In order
  to ignore complications arising from these scratches in the analysis of the
  data we have chosen to only analyze the unblemished region of the array.
\end{thebibliography}
\end{document}